\documentclass[aps,prl,twocolumn,superscriptaddress]{revtex4-2}

\usepackage{makeidx,amssymb,amsmath,amsthm,graphicx}
\usepackage{lineno}

\usepackage{xcolor}

\begin{document}

\title{Evolution of electron spin resonance\\ through a metallic quantum critical phase diagram}


\author{Marc Scheffler}
\thanks{Both authors contributed equally.}
\affiliation{1. Physikalisches Institut, Universit\"at Stuttgart, 70569 Stuttgart, Germany}
\author{J\"org Sichelschmidt}
\thanks{Both authors contributed equally.}
\affiliation{Max-Planck-Institut f\"ur Chemische Physik fester Stoffe, 01187 Dresden, Germany}
\author{Conrad Clauss}
\affiliation{1. Physikalisches Institut, Universit\"at Stuttgart, 70569 Stuttgart, Germany}
\author{Mojtaba Javaheri Rahim}
\affiliation{1. Physikalisches Institut, Universit\"at Stuttgart, 70569 Stuttgart, Germany}
\author{Boris I. Kochelaev}
\thanks{Deceased.}
\affiliation{Max-Planck-Institut f\"ur Chemische Physik fester Stoffe, 01187 Dresden, Germany}
\author{Cornelius Krellner}
\affiliation{Max-Planck-Institut f\"ur Chemische Physik fester Stoffe, 01187 Dresden, Germany}
\affiliation{Physikalisches Institut, Goethe-Universit\"at Frankfurt, 60438 Frankfurt/Main, Germany}
\author{Christoph Geibel}
\affiliation{Max-Planck-Institut f\"ur Chemische Physik fester Stoffe, 01187 Dresden, Germany}
\author{Frank Steglich}
\affiliation{Max-Planck-Institut f\"ur Chemische Physik fester Stoffe, 01187 Dresden, Germany}
\affiliation{Center for Correlated Matter and School of Physics, Zhejiang University, Hangzhou 310058, China}
\author{Martin Dressel}
\affiliation{1. Physikalisches Institut, Universit\"at Stuttgart, 70569 Stuttgart, Germany}


\date{\today}

\begin{abstract}
In the heavy-fermion metal YbRh$_{2}$Si$_{2}$, quantum criticality at a suppressed antiferromagnetic order is governed by the interplay of local magnetic moments and itinerant conduction electrons. 
We demonstrate how this can be investigated by a new experimental approach that enables the observation of electron spin resonance (ESR) across a broad range of frequencies and fields at very low temperatures. This allowed us to cover a large part of the phase diagram from the paramagnetic Fermi-liquid phase to the phase with antiferromagnetic order and including the quantum-critical regime. Both the ESR $g$ factor and the linewidth present distinct behaviors in these three regimes, providing further insight into the physics across a quantum critical point. 
Notably, when cooling down at a field directly towards the quantum critical point, both $g$ factor and linewidth continuously decrease. Furthermore, we observe a very good matching of the $g$ factor behavior upon field-tuning and temperature-tuning towards the quantum-critical point. We analyze and discuss the results in the context of present theories on ESR in strongly correlated electron systems. 
\end{abstract}


\maketitle

Heavy-fermion metals such as YbRh$_{2}$Si$_{2}$ are governed by strong electronic interactions. In particular, YbRh$_{2}$Si$_{2}$ is a realization of a Kondo lattice where the unfilled $f$-shell of each Yb atom in the crystal lattice results in a local magnetic moment, which interacts with conduction electrons. At low temperatures, this so-called Kondo interaction at the Yb sites becomes coherent, local moments and conduction electrons hybridize and a new metallic heavy-fermion state appears, governed by mobile electrons of strongly enhanced effective mass. This is a non-magnetic, paramagnetic ground state as the $f$-derived local moments are Kondo-screened by the conduction electrons. If instead the intersite interaction between the moments is dominant, in many cases an antiferromagnetic (AF) ground state appears. The tuning between both ground state regimes, at zero temperature, is a quantum phase transition, which can lead to distinct and unconventional properties at finite temperatures caused by quantum criticality \cite{gegenwart08a,si10b}.\\

\begin{figure}[bt]
\includegraphics[width=1\columnwidth]{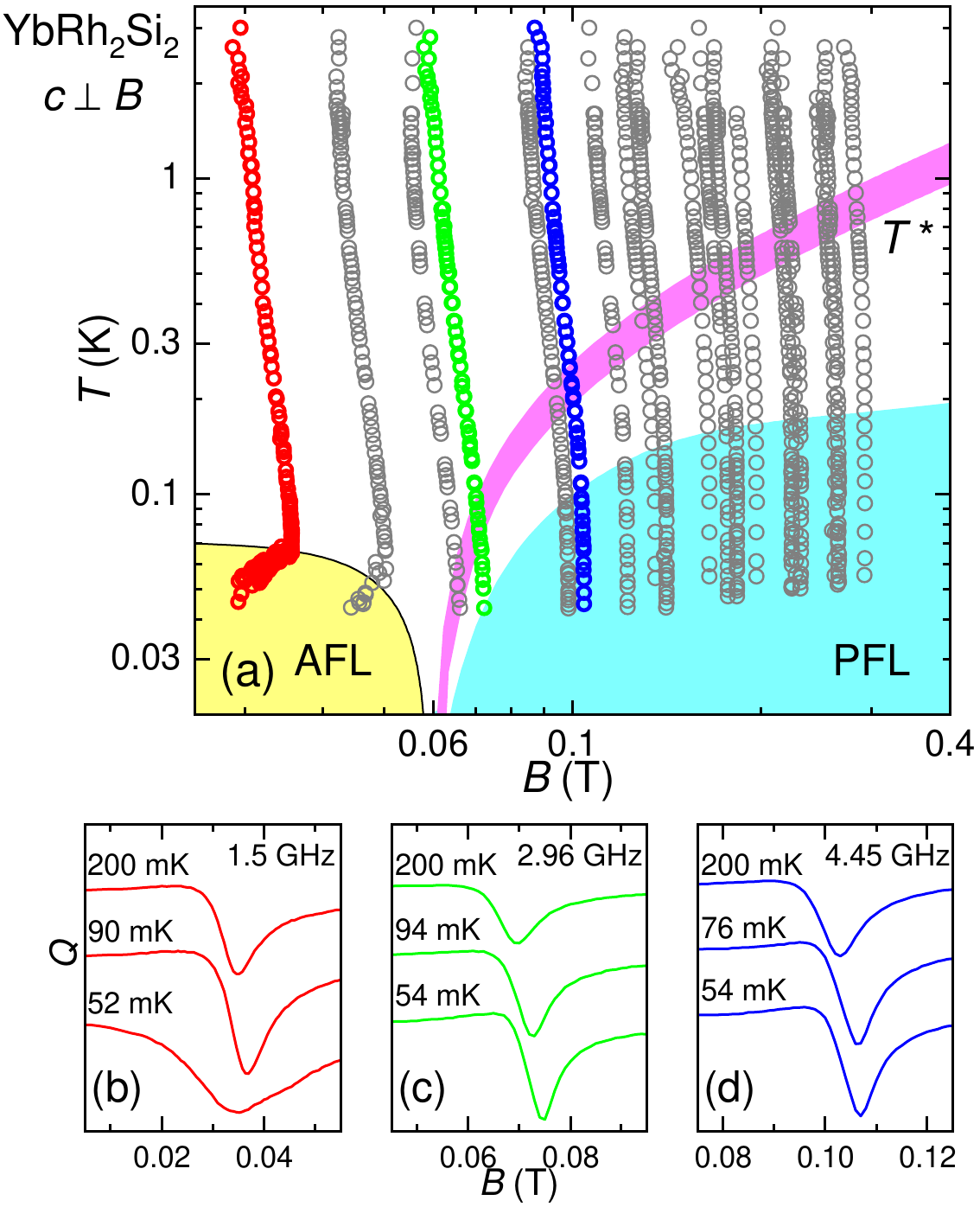}
\caption{Probing the phase diagram of YbRh$_{2}$Si$_{2}$ with ESR. (a) Overall phase diagram with extremely heavy LFL regimes showing AF order (AFL) and metallic paramagnetism (PFL), as well as the $T^{*}$ line \cite{gegenwart08a}. The circles indicate the ESR resonance field for different microwave frequencies obtained by temperature-dependent ESR measurements, i.e.\ we probe all relevant regimes. The pronounced evolution of these resonance fields as function of temperature clearly differs for the different regimes, and directly reflects the temperature-dependent $g$ factor. (b, c, d) Exemplary ESR spectra (background-corrected resonator quality factor $Q$ as function of magnetic field $B$) for three resonance frequencies, with matching colors in (a), for three temperatures each. Key observations upon cooling are (b) broadening and back-shift in the AFL phase, (c) monotonous shift and narrowing in the quantum-cricital regime, and (d) temperature-independent resonance field in the PFL regime. Spectra for further temperatures are presented in the Supplemental Materials \cite{SM1}.
}
\label{Fig1}
\end{figure}

YbRh$_{2}$Si$_{2}$ is a prime example of such a quantum-critical heavy-fermion system. When cooled in absence of a magnetic field $B$, YbRh$_{2}$Si$_{2}$ orders antiferromagnetically below the very low N\'{e}el temperature $T_{\rm N}= 70$~mK \cite{trovarelli00a}. This magnetic order can be suppressed if a moderate magnetic field $B$, exceeding the critical field of 60 mT, is applied perpendicular to the $c$-axis of the tetragonal crystal structure. The phase diagram is displayed in Fig.~\ref{Fig1}(a) where at $T=0$ the quantum-critical point separates the 
AF order below $T_{\rm N}(B)$ from a paramagnetic (PM) metallic state at elevated fields. Both states are extremely heavy Landau Fermi liquids \cite{gegenwart08a}, in the following labeled ‘AFL’ and ‘PFL’, respectively. 
The Landau Fermi liquid (LFL) is a canonical model for conventional metals where electronic interactions are renormalized according to the quasiparticle concept that interprets the metallic many-body state within a single-electron picture. At finite $T$, there is a wide range in the phase diagram with unconventional metallic properties caused by the presence of the quantum-critical point.
This quantum-critical regime features numerous properties that are at odds with LFL theory, and dubbed non-Fermi liquid (NFL). The phase diagram of YbRh$_{2}$Si$_{2}$ close to this quantum-critical point has been studied in great detail with magnetic, thermodynamic, and transport measurements, and these have not only established the canonical quantum-critical phase diagram but also an additional temperature scale \cite{paschen04a,gegenwart07a} which increases with magnetic field. This $T^{*}$ is considered a manifestation of the crossover from the low-field state, where the $f$-electrons are of local nature, to a state where they are delocalized and contribute to composite charge carriers. The exact nature, however, is not completely understood \cite{kummer15a,generalov17b,jang20a,paschen16a}, alternative pictures were discussed \cite{wolfle15a}, and recent experimental results \cite{schubert19a} support the notion that $T^{*}$ is associated with a partial ferromagnetic (FM) polarization of fluctuating magnetic moments \cite{gegenwart05a}.
Moreover, a distinct inelastic scattering process of the heat carriers at $T^{*}$ \cite{pfau12a,pfau13a} may be considered a finite-temperature precursor to a violation of the Wiedemann-Franz law at the quantum critical point \cite{pfau12a,steglich14a}
\footnote{A different conclusion is drawn by Taupin et al., Phys. Rev. Lett. \textbf{115}, 046402 (2015), cf. also \cite{schuberth22a}.}.
The nature of the AFL phase remains enigmatic: On the one hand a LFL phase forms upon cooling through $T_{\rm N}(B)$ \cite{gegenwart02a,custers03a,paschen04a,pfau12a}, with effective quasiparticle masses even considerably larger than those in the PFL phase \cite{gegenwart08a}, while on the other hand, AF ordering sets in with staggered moments possibly along the hard crystal-electric-field direction \cite{hafner19a,hamann19a} as tiny as $2\times10^{-3}\mu_{\rm B}$/Yb \cite{ishida03a}. Further on, AF ‘hybrid nuclear, $4f$-electronic’ order \cite{schuberth16a} resp. ‘electro-nuclear’, spatially modulated \cite{knapp23a} order onsets at low mK temperatures and is intimately connected with the remarkable formation of a heavy-fermion superconducting phase which is currently considered to be based on a spin triplet pair density wave \cite{schuberth16a,nguyen21a,schuberth22a,knapp23a,levitin26a}.\\ 

Spectroscopy is the desired experimental approach to address the different energy and temperature scales of YbRh$_{2}$Si$_{2}$. Unfortunately, the most common spectroscopic techniques such as neutron scattering, scanning tunneling, photoemission, optical spectroscopy, or nuclear magnetic resonance face severe technical limitations for the particular case of YbRh$_{2}$Si$_{2}$ \cite{kummer15a,kimura06a,stock12a,ernst11a,kambe14a,usachov24b}.
We will show that electron spin resonance (ESR) emerges as a novel spectroscopic tool that is ideally suited for the study of YbRh$_{2}$Si$_{2}$ close to the quantum-critical point.

ESR is a powerful local probe to address intrinsic magnetic properties. If in the presence of a static magnetic field a transverse microwave magnetic field of frequency $\nu$ matches the resonance condition $h\nu_{\rm res}=g\mu_{\rm B}B_{\rm res}$ for the $g$ factor of the material under study (with $h$ Planck's constant and $\mu_{\rm B}$ Bohr magneton), microwave power is absorbed. The two most important parameters to describe ESR are $g$, being a property of the resonant magnetic moment, and the ESR linewidth $\Delta B$, which for metals generally corresponds to the transverse spin relaxation rate $1/T_{2}$ that is governed by the spin dynamics of the surrounding \cite{barnes81a,elschner97a}.
The local moments are screened by Kondo interaction, and commonly the magnetic moments in heavy-fermion metals are not accessible to ESR techniques.
Therefore, it came as a great surprise when a pronounced and narrow ESR line was detected in YbRh$_{2}$Si$_{2}$ \cite{sichelschmidt03a} and a few other heavy-fermion metals with pronounced FM fluctuations \cite{krellner08a,hafner19a}. YbRh$_{2}$Si$_{2}$ and chemically substituted variants now serve as model systems for ESR in Kondo lattices \cite{duque09a,sichelschmidt10b,gruner12a,huber12a,kochelaev17a}.

The theoretical understanding developed twofold. A local approach found that, while on-site Kondo screening causes an immeasurably huge linewidth, coherent Kondo scattering in a Kondo-lattice system crucially supports a narrow, measurable ESR signal \cite{kochelaev17a}. 
This is mainly due to the formation of a collective spin mode of quasi-localized $f$-electrons and wide-band conduction electrons even in the presence of strong anisotropies in the Kondo interaction \cite{kochelaev09a}. An itinerant approach could qualitatively explain the results by a heavy quasiparticle resonance \cite{wolfle09a}. Both approaches account for FM correlations, providing an important additional contribution for the narrowing of the resonance lines.

Since conventional ESR spectrometers operate at a fixed microwave frequency and scan the magnetic field, probing the different phases of YbRh$_{2}$Si$_{2}$ at certain magnetic fields requires different ESR frequencies \cite{schaufuss09a,duque09a,sichelschmidt10a}.
Most of these experiments were performed at ESR fields of $\approx 200$~mT and higher, and all of them were performed at temperatures above 400 mK. This implies that the most interesting region of the phase diagram, at $T<100$~mK and $B<100$~mT, was not accessible. Here we overcome previous limitations by introducing a new experimental approach that operates at low-GHz frequencies, being sensitive enough for YbRh$_{2}$Si$_{2}$, and compatible with mK temperatures in a dilution refrigerator.

We employ superconducting planar resonators based on a one-dimensional transmission line \cite{scheffler13a} that can easily be operated at several harmonics. Using meander shaped lines, resonator frequencies as low as 1.5 GHz are implemented on a chip with size of just a few mm. Resonators of different length enable us to vary the ESR frequency as desired. Using three resonators at $40~{\rm mK} < T <4$~K, we have performed ESR measurements on YbRh$_{2}$Si$_{2}$ in a wide range of the low-temperature phase diagram, as illustrated 
in Fig.~\ref{Fig1}(a).
The data acquisition and line-shape analysis are described in the Supplemental Materials \cite{SM1}.
In the following presentation of our ESR results we start with the low-field AFL region, continue with the PM region at elevated fields, and finally consider the quantum critical regime between AFL and PFL states.

\begin{figure}[b]
\includegraphics[width=1\columnwidth]{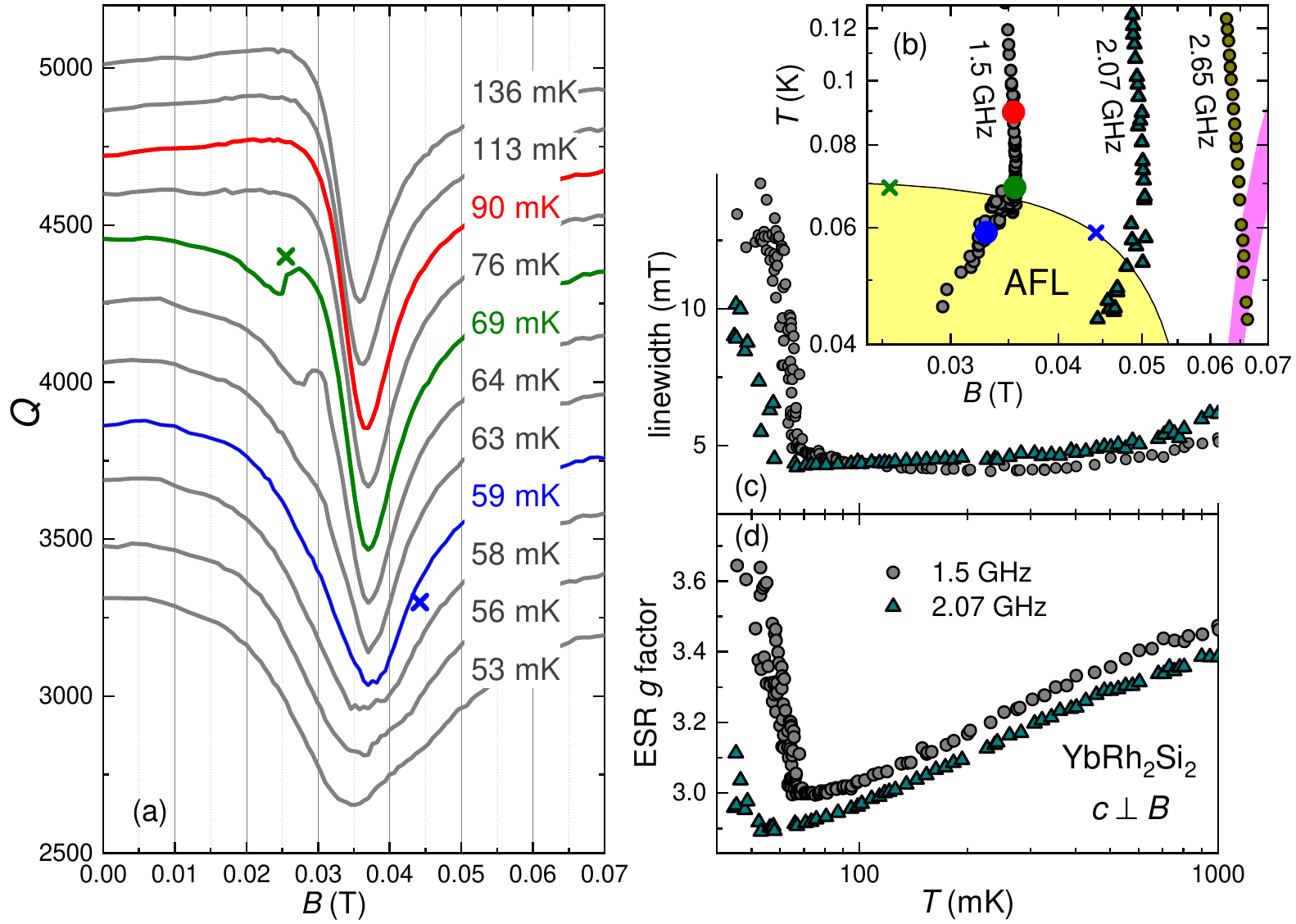}
\caption{ESR of YbRh$_{2}$Si$_{2}$ in the AFL phase and crossing the AF transition. (a)~ESR spectra for selected temperatures at 1.5 GHz, including temperature and field ranges in the AFL phase. $Q$ axis refers to 53 mK data, higher temperature data are shifted by 200 each. (b)~ESR resonance fields for the three lowest resonator frequencies at low temperatures. Upon cooling, the modes at 1.5 and 2.07 GHz change drastically when entering the AFL phase. Data points at 59 mK, 69 mK, and 90 mK of the 1.5 GHz resonator correspond to colored spectra in (a). The crosses in (a) and (b) correspond to fields of the AFM transition at 59 and 69 mK. (c)~ESR linewidth $\Delta B$ and (d) $g$ factor increase drastically in the AFL phase for the two lowest frequencies.}
\label{Fig2}
\end{figure}
From Fig.~\ref{Fig1}(a) one can deduce the general temperature dependence of the ESR resonance field for a given resonator frequency within the phase diagram of YbRh$_{2}$Si$_{2}$, and this also becomes apparent already from the few exemplary ESR spectra in Fig.~\ref{Fig1}(b-d). Upon cooling the ESR typically shifts to higher fields and becomes more pronounced, consistent with previous measurements at higher $T$ \cite{sichelschmidt03a,scheffler13a}. However, this general trend does not hold at low temperatures. 
Whereas in the PFL regime the ESR does not change much upon cooling further, the effects on the ESR are dramatic when entering the AFL phase, despite the very weak magnetic moments of the AFL phase \cite{ishida03a}.

The ESR in and near the AFL phase is illustrated in detail in Fig.~\ref{Fig2} for the two resonators with 1.5 and 2.07 GHz: the resonance fields strongly decrease, $g(T)$ correspondingly increase (Figs.~\ref{Fig2}(b) and (d)) and the ESR line becomes much broader than in the PM phase (Fig.~\ref{Fig2}(c)). 
Noteworthy, this evolution of the resonance field is the opposite behaviour of the ordered resonance modes in CeRuPO \cite{forster10b} and Yb(Rh$_{\rm 0.73}$Co$_{\rm 0.27}$)$_2$Si$_2$ \cite{fazlizhanov26a} which both display FM order along the $c$-axis being the hard crystalline electric field direction.  
In the region closely above the AFL phase $\Delta B(T)$ tends to saturate or to reach a minimum around 0.3~K while $g(T)$ is continuously decreasing, see Figs.~\ref{Fig2}(c) and (d). The linewidth behavior indicates less effective narrowing due to slowing down of magnetic fluctuations or a reduced effect of FM correlations \cite{taylor75b, krellner08a}. The temperature and field dependence of the $g$ factor reflect the evolution of the internal field above the ordered state. As detailed in \cite{SM1} we could describe this by a molecular field model assuming FM exchange in the plane and AF interaction between the planes. Within this model the resulting $g$ factor in the ordered phase is expected to reach $g_{\rm eff}^{\rm AF}\approx3.72$. This is considerably larger than the observed $g_{\rm exp}^{\rm AF}\approx3.4$, which is probably due to Kondo screening.   

ESR spectra in the field region of the AFL phase are presented in Fig.~\ref{Fig2}(a) for selected temperatures around $T_{\rm N}(B)$. Their structure is expected to depend on the interplay between internal and external magnetic fields, thus containing information on the magnetically ordered structure as was analyzed for the well-localized magnetic moments in the ordered state of GdRh$_{2}$Si$_{2}$ \cite{sichelschmidt18a}. However, the analysis for YbRh$_{2}$Si$_{2}$ is less clear and the spectra contain structures merely related to the phase boundary $T_{\rm N}(B)$ as illustrated in Fig.~\ref{Fig2}(a).
The spectra at $T=69$ and 64~mK, just below $T_{\rm N}(B = 0)$, exhibit a sharp kink at fields of 25 and 27~mT, respectively. These fields correspond to the critical AF field, as indicated by the green crosses in Figs.~2(a) and (b). These kinks can be understood as the result of the transition from a very broad line in the AFL regime at low fields to a narrow line in the PM regime at higher fields. Because at these temperatures the transition occurs at fields well below the resonance field, at the transition the signal of the resonance is already sizable for the broad AFL line, but much weaker for the narrow line of the PM state, resulting in an upward step at the transition AFL - PM. There seem to be similar features on the high field side at about 45~mT in the $T\le59$~mK spectra which match also quite well with the $T_{\rm N}(B)$ curve (blue crosses in Figs.\ref{Fig2}(a) and (b)). However, these features are much weaker, preventing a strong statement.  We note that the simultaneous observation of an AFL and a PM-type ESR in the same spectra does not mean a coexistence of the two phases, since the two signals are observed in distinct field ranges.

\begin{figure}[htbp]
\includegraphics[width=1\columnwidth]{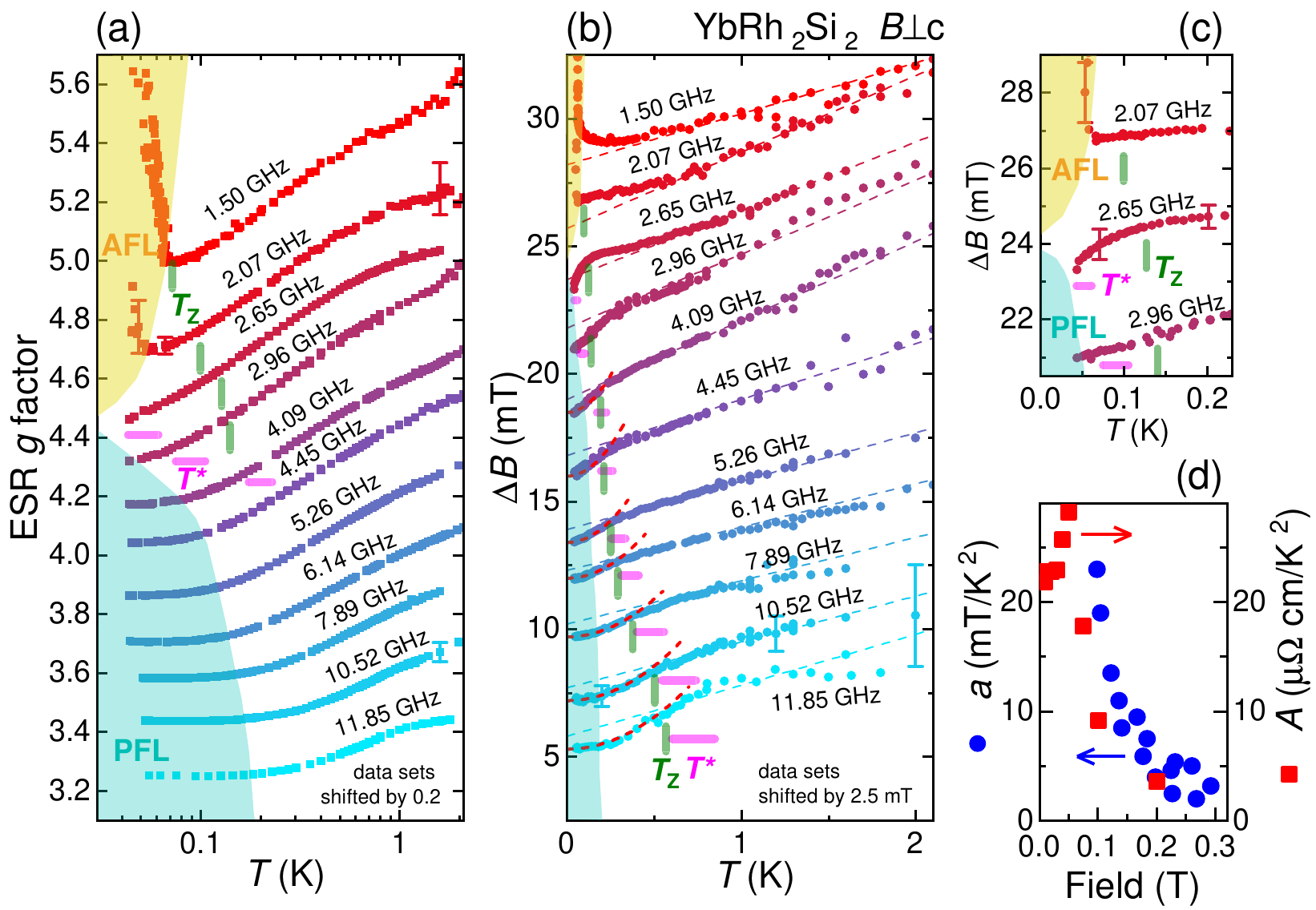}
\caption{Temperature dependence of ESR parameters, with focus on paramagnetic regimes. (a) ESR $g$ factor and (b) linewidth $\Delta B$ for exemplary resonator frequencies. The transition/crossover temperatures for antiferromagnetic (AFL) / paramagnetic (PFL) regimes are indicated by yellow and cyan areas, respectively; the $T^{*}$ crossover range is shown by magenta bars, and the temperature $T_{\rm Z}$ that corresponds to the Zeeman energy ($k_{\rm B} T_{\rm Z} \equiv h\nu_{\rm res}=g\mu_{\rm B}B_{\rm res}$) is indicated by vertical lines for magnetic fields close to the critical field. This region is enlarged in panel (c). Straight dashed lines are guides to the eye; parabolic dashed lines represent ($aT^2$) with field-dependent coefficients $a$ plotted in (d):  Comparison of $a$ to the quasiparticle scattering cross section $A\propto m^{\ast2}$ determined from electrical resistivity \cite{gegenwart02a}.  }
\label{Fig3}
\end{figure}
Next we turn to the ESR results in the PM region which contains the $T^{*}$ and PFL regimes. The PFL state of YbRh$_{2}$Si$_{2}$ is experimentally well established, in particular the $T^{2}$-behavior of the dc resistivity served as indicator for the PFL temperature $T_{\rm PFL}$ as a function of field in the phase diagram of Fig.~\ref{Fig1}(a) \cite{gegenwart02a}.
Note, however, that this is not a sharp transition but a crossover between two metallic regimes.
As far as the ESR behavior in the PFL regime is concerned, in general one expects a quadratic temperature dependence of the linewidth while the $g$ factor is temperature independent \cite{wolfle09a}; however, this was never directly observed.
In Fig.~\ref{Fig3}(a) we clearly see the change from a strongly temperature-dependent $g$ factor in the NFL regime above $T_{\rm PFL}$ to a $T$-independent $g$ factor below.
Utilizing the numerous ESR frequencies available to us,
we can track the temperature evolution of the $T_{\rm PFL}$ line that fully reproduces the established $T_{\rm PFL}(B)$ dependence.

From Fig.~\ref{Fig3}(b) we see that the ESR linewidth exhibits a distinct s-shaped temperature dependence in the PM range. Starting with the flattened part in the PFL regime, for $B>100$~mT ($\nu_{\rm res}>4.45$~GHz) the lines significantly broaden above $T_{\rm PFL}$ until $\Delta B(T)$ appears to change around $T^{*}$, approaching the known linear dependence for $T > T^{*}$ \cite{schaufuss09a,sichelschmidt10a}.
Our detailed low-temperature data now allow a much closer look for fields toward the quantum critical field. There the suppression of $\Delta B(T)$ upon cooling becomes pronounced already at the temperature of the Zeeman energy, $k_{\rm B} T_{\rm Z}\equiv g\mu_{\rm B}B_{\rm res} = h\nu_{\rm res}$ rather than at $T^{*}$, becoming most pronounced near the quantum phase transition as can be well seen for the 2.65 GHz data enlarged in Fig.~\ref{Fig3}(c). This behavior is in agreement with the critical slowing-down seen by THz spectroscopy \cite{yang23a} and is consistent with the expected $T^{3/4}$-behavior for spin-flip scattering of critical quasiparticles \cite{wolfle15a}.
In the regime $B>100$~mT and $T< T^{*}$, the linewidth can be well described by $\Delta B(T) = a T^2 +{\rm const.}$ as indicated by the dashed lines in Fig. \ref{Fig3}(b).
The field dependence of the prefactor $a$ resembles a similar behavior of the corresponding factor $A$ in the dc resistivity $\rho(T) = AT^2 +{\rm const.}$  \cite{gegenwart02a} observed in the PFL regime [Fig. \ref{Fig3}(d)]. Hence, the signatures of PFL are visible in $\Delta B(T)$ up to $T^{*} > T_{\rm PFL}$. 
This could be due to probing the PFL properties by the linewidth being a dynamic quantity in contrast to the static quantities $\rho(T)$ and $g(T)$.
An additional narrowing of the lines below $T^{*}$ can qualitatively be understood by FM correlations stabilizing the resonance \cite{gegenwart05a,krellner08a}. Noteworthy, in contrast to the ac-susceptibility \cite{gegenwart07a}, no maximum occurs in the $g$ factor around $T^{*}$ \cite{notegmax}.\\
\begin{figure}
\includegraphics[width=1\columnwidth]{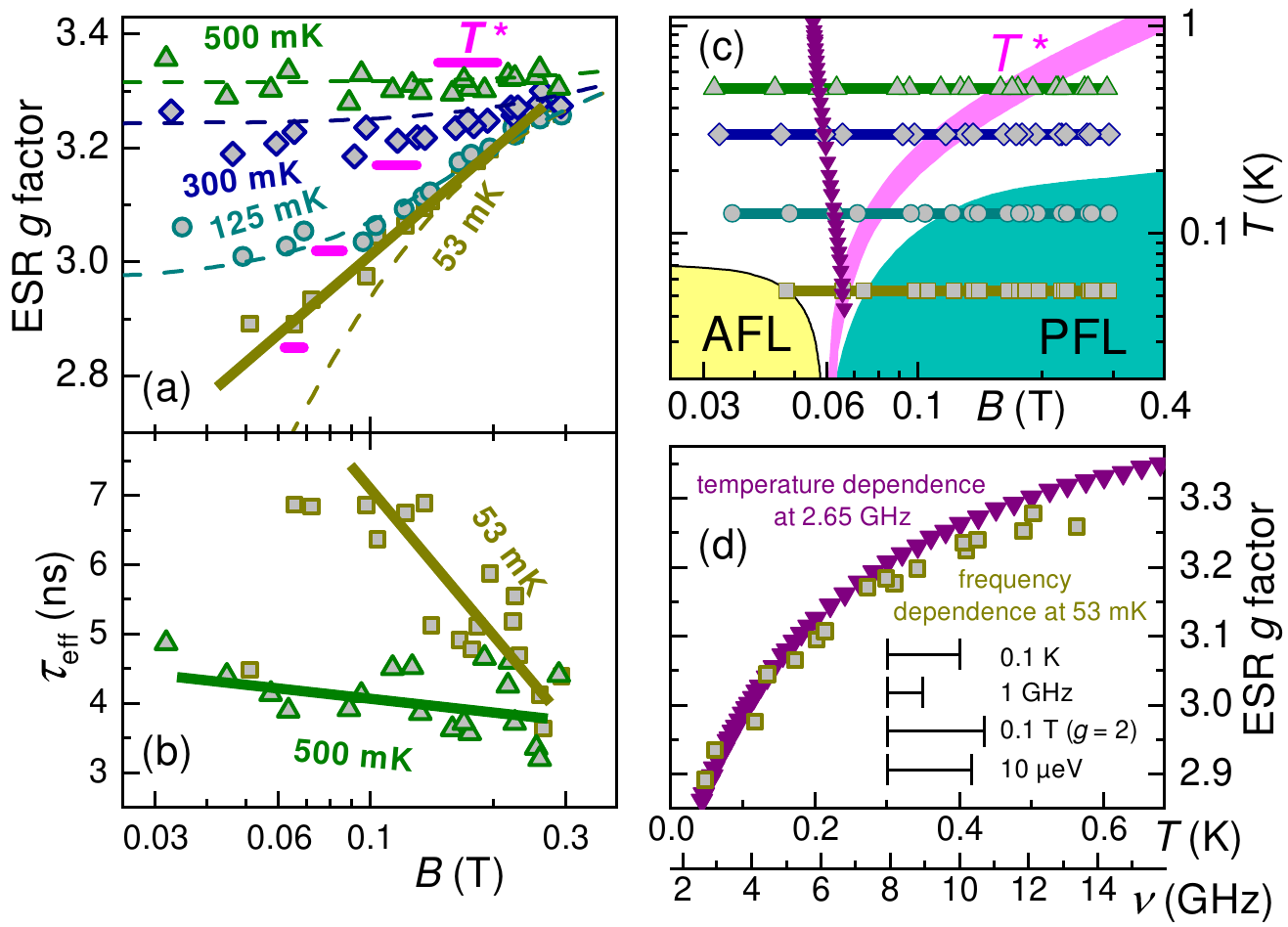}
\caption{Magnetic field dependence of ESR parameters in the paramagnetic regimes. (a) $g$ factor and (b) effective ESR relaxation time $\tau_{\rm eff}= B_{\rm res}/\nu\Delta B$, for temperatures and resonance fields as indicated in panel (c).
Solid lines guide the eye; dashed lines: molecular field model \cite{SM1}.
(c) Phase diagram of YbRh$_{2}$Si$_{2}$ with horizontal-line connected symbols (resonance fields) referring to data shown in the other frames. The solid triangles denote the resonance fields of the 2.65~GHz mode.
(d) Comparison of the $g$ factor temperature dependence in the quantum-critical regime (horizontal axis: $T$~(K); $\nu =$ 2.65~GHz mode) and frequency dependence in the PFL regime (horizontal axis: $\nu$~(GHz); $T \approx$ 53~mK). The two horizontal axes as well as additional ranges in the figure indicate the quantitative relation of different relevant energy scales.
To achieve matching of the two data sets in (d), the frequency axis is shifted by 1.68~GHz, which corresponds to the ESR frequency for $g=2$ at the critical field of 60~mT.
}
\label{Fig4}
\end{figure}
Finally let us consider the quantum-critical regime [Fig.~\ref{Fig4}(c)], which only now is fully accessible for ESR measurements in temperature and field.
In particular the 2.65 GHz data track the $T$ dependence in the NFL regime over two orders of magnitude. The continuous evolution in the $g$ factor evidences a logarithmic behavior in a wide temperature range. 
Such logarithmic dependence early-on indicated that a single-ion Kondo scenario was applicable \cite{sichelschmidt03a}. Later it was shown that the Kondo effect supports the ESR signal by a collective spin mode of Yb spins and conduction electron spins \cite{kochelaev09a}.
The same behavior in $g(T)$ is found for all other modes in the NFL regime.

While the field dependence of the ESR parameters, displayed in Figs.~\ref{Fig4}(a) and (b),
is rather weak above 500 mK, the behavior changes drastically at lower temperatures.
For $T=125$~mK, the $g$ factor clearly increases with magnetic field, even more pronounced for fields above the $T^{*}$ scale.
At even lower temperatures, $T= 53$~mK, we observe a drastic rise of $g$ for fields above the critical field \cite{SM1}.

Starting from thermodynamic data, Fermi liquid theory allows a renormalization of the quasiparticle Zeeman splitting \cite{wolfle15a}. As discussed in \cite{SM1} (Fig.\ S7), these theoretical estimates yield much larger values and show much stronger field dependencies than the renormalization extracted from the measured $g$ factor, which is a direct spectroscopic experiment. The discrepancies may be a result of the dual itinerant-localized character of the $f$ electrons in such strongly correlated $f$-electron systems.

At the lowest temperature of 53 mK, the ESR relaxation time $\tau_{\rm eff}$ increases strongly when approaching the quantum-critical field. This is a signature of the coefficient $a(B)$ shown in Fig.\ \ref{Fig3}(d), and reflects the field dependence of the quasiparticle scattering cross section extracted from the dc resistivity \cite{sichelschmidt10a}.
This picture is consistent with the scenario of a fermionic critical slowing down near a magnetic quantum phase transition \cite{yang23a}.\\
The low-temperature $g$ factor increases logarithmically with field well into the PFL regime;
this reminds of the similar $T$ dependence in the NFL range.
For direct comparison, we make two cuts through the phase diagram in Fig. \ref{Fig4}(c):
The vertical cut yields $g(T)$ measured at a fixed frequency of 2.65 GHz, i.e. very close to the critical field. The horizontal cut gives the $g(B)$ for the lowest temperature of $T=53$~mK.
Fig.\ \ref{Fig4}(d) presents the energy dependence of the $g$ factor
for $h\nu$ and $k_{\rm B}T$.
Since the quantum critical point at $T = 0$ and $B = 60$~mT is the conceptual starting point
for both dependencies, the origin of the frequency axis has to be shifted by 1.68 GHz.
The overlaps of the two data sets is extraordinary and suggests an intrinsic relation between the frequency/field dependence in the PFL regime and the temperature dependence in the quantum critical regime. We propose to look for the same correspondence in other physical quantities, e.g.\ the specific heat.
The necessary frequency shift of 1.68 GHz corresponds to the ESR frequency at 60 mT for $g = 2$.
This could imply that the ESR signal in YbRh$_{2}$Si$_{2}$ evolves from free-electron-like ($g = 2$) at the QCP to local Yb moments ($g = 3.777$ in this tetragonal crystal environment \cite{kutuzov08a}) at high temperatures and high fields.
Clearly, much lower temperatures and precise control of the ESR frequency \cite{wiemann15a} would be required to check whether the ESR signal really converges to $g=2$.
The $g$ factor approaching $g = 2$ right at the QCP can be regarded the first experimental verification of the theoretically proposed "loss of (composite heavy) quasiparticles at a Kondo-breakdown QCP" \cite{hu24a}. It proves a unique view through ESR into Mott-type quantum criticality, as realized in YbRh$_{2}$Si$_{2}$.

Tracking the ESR evolution towards the QCP (at the critical field \cite{wiemann15a} and at lower temperatures) might help to disentangle the role of the static magnetic field as a tuning parameter for both the phase diagram of YbRh$_{2}$Si$_{2}$ and the ESR frequency.
The increase of the linewidth upon entering the AFL phase could be caused not only by a common inhomogeneous broadening effect but also by the notion that resonance-stabilizing FM fluctuations \cite{krellner08a} are not relevant in this phase as can be concluded, e.g., from early NMR investigations in the PM region close to the QCP \cite{ishida02a}. 
Overall, our new experimental approach can help to unravel the enigmatic nature of the ordered phase and is able to fully cover the interesting regimes of quantum criticality (NFL) and paramagnetic LFL (PFL). Therefore, it should also be applied to other quantum-critical heavy-fermion materials where an ESR signal was observed, such as $\beta$-YbAlB$_{4}$ \cite{holanda11a}, or to other spin systems with novel low-energy phases \cite{pratt11a}.

\begin{acknowledgments}
We remember Prof. Boris I. Kochelaev, who sadly passed away shortly before this manuscript was completed \cite{aganov25a}. 
We thank M. Brando, S. Kirchner, H.-A. Krug von Nidda, Q. Si, and P. W\"olfle for helpful discussions and D. Bothner, D. K\"olle, and R. Kleiner for support in resonator preparation. We are grateful for detailed remarks by D. Bothner. We acknowledge financial support by the DFG through Grants No. SCHE 1580/2-1, SI 1339/1-1, 
KR3831/4-1, via the TRR 21 (project A3), and via the TRR 288 (422213477, project A03) and the Volkswagen Stiftung (I/84689).
\end{acknowledgments}
\bibliography{YRSmKESRfinal}

\end{document}


\title{Supplementary Material of ``Evolution of electron spin resonance\\ through a metallic quantum critical phase diagram''}
\author{Marc Scheffler}
\thanks{Both authors contributed equally.}
\affiliation{1. Physikalisches Institut, Universit\"at Stuttgart, 70569 Stuttgart, Germany}
\author{J\"org Sichelschmidt}
\thanks{Both authors contributed equally.}
\affiliation{Max-Planck-Institut f\"ur Chemische Physik fester Stoffe, 01187 Dresden, Germany}
\author{Conrad Clauss}
\affiliation{1. Physikalisches Institut, Universit\"at Stuttgart, 70569 Stuttgart, Germany}
\author{Mojtaba Javaheri Rahim}
\affiliation{1. Physikalisches Institut, Universit\"at Stuttgart, 70569 Stuttgart, Germany}
\author{Boris I. Kochelaev}
\thanks{Deceased.}
\affiliation{Max-Planck-Institut f\"ur Chemische Physik fester Stoffe, 01187 Dresden, Germany}
\author{Cornelius Krellner}
\affiliation{Max-Planck-Institut f\"ur Chemische Physik fester Stoffe, 01187 Dresden, Germany}
\affiliation{Physikalisches Institut, Goethe-Universit\"at Frankfurt, 60438 Frankfurt/Main, Germany}
\author{Christoph Geibel}
\affiliation{Max-Planck-Institut f\"ur Chemische Physik fester Stoffe, 01187 Dresden, Germany}
\author{Frank Steglich}
\affiliation{Max-Planck-Institut f\"ur Chemische Physik fester Stoffe, 01187 Dresden, Germany}
\affiliation{Center for Correlated Matter and School of Physics, Zhejiang University, Hangzhou 310058, China}
\author{Martin Dressel}
\affiliation{1. Physikalisches Institut, Universit\"at Stuttgart, 70569 Stuttgart, Germany}

\maketitle

\section{Resonator fabrication and sample fabrication}
The superconducting coplanar resonators were prepared by sputtering a 150 nm thick Nb film onto $430\, \mu\rm m$ thick $r$-cut sapphire substrate, conventional UV lithography of the coplanar resonator structure, and dry etching. The resonator design is similar to established layouts from quantum circuitry \cite{goeppl08a, kubo10a, bothner12a}, with the coplanar resonator waveguide in meander shape to achieve low fundamental frequencies with compact design and substantial coverage of the resonator by the sample to be probed by ESR \cite{scheffler13a,javaherirahim16a,bondorf18a,miksch21a}. We have obtained similar results using superconducting stripline resonators \cite{scheffler13a}. Single-crystalline YbRh$_{2}$Si$_{2}$ samples were prepared as those of precedent ESR experiments \cite{sichelschmidt10b}, as described previously \cite{krellner12a}. The presented data were obtained on a platelet-shaped sample (batch 63114, RRR$\approx$50, see also Supplement of Ref.\ \cite{schuberth16a}) with lateral dimension (in the $ab$-plane) of approximately 2~mm.

\renewcommand{\thefigure}{S1}
\begin{figure}[tbp]
\begin{center}
\includegraphics[width=0.9\columnwidth]{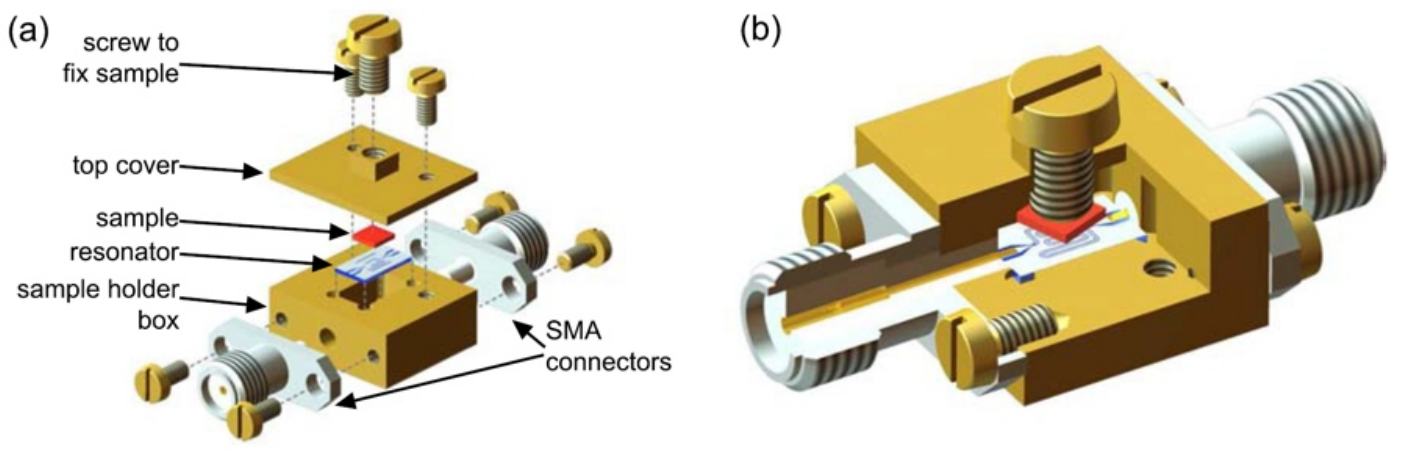}
\caption{
Design of the microwave setup with resonator and sample mounting, shown as (a) Exploded-view drawing of the assembly and (b) cutaway drawing indicating the actual positioning of the sample with respect to the resonator. The superconducting resonator, with main straight sections of the coplanar meander geometry aligned parallel to the static magnetic field, is mounted in a gold-plated brass box and connected to coaxial SMA connectors. The top cover lid of the brass box holds a screw, to which the YbRh$_{2}$Si$_{2}$ sample is attached \cite{ConradClaussDissertation}.
}
\label{FigSetup}
\end{center}
\end{figure}
\section{Microwave measurements, sample mounting, and temperature determination}
For the microwave measurements, a resonator assembly was mounted on a coldfinger attached to the coldplate of an Oxford Instruments Kelvinox400 dilution refrigerator, which is equipped with a superconducting solenoid magnet for fields up to 8~T. The sample was attached to a brass screw, which holds the sample at a certain distance above the resonator chip, see Fig.~\ref{FigSetup}.
This brass screw was thermally coupled to the mixing chamber coldplate by a 1~mm thick silver wire. The sample temperature was monitored by a ruthenium oxide temperature sensor attached to the silver wire as close to the sample as possible.
Static magnetic fields were applied parallel to the ab-plane of the YbRh$_{2}$Si$_{2}$ sample and the resonator chip.
The microwave signal transmitted through the resonator was measured with a vector network analyzer (VNA, Agilent E8364C). 
Extra efforts were taken to make sure that the measured sample temperature is indeed the temperature of the active sample volume. The latter is the skin depth region, approximately a few $\mu\rm m$ thick, on the side of the sample facing the resonator, while the temperature is measured on the back side of the approximately 0.3 mm thick sample. 
\renewcommand{\thefigure}{S2}
\begin{figure}[tbp]
\begin{center}
\includegraphics[width=0.85\columnwidth]{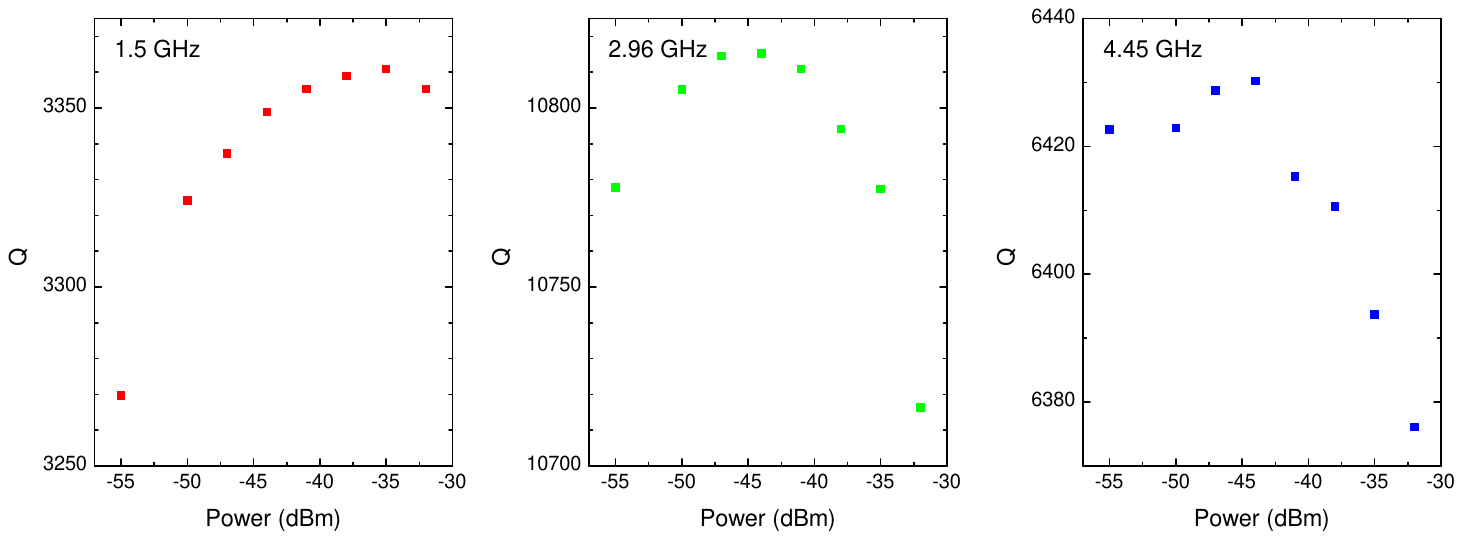}
\caption{
Exemplary power dependencies of the quality factor for the first three modes of the resonator R01 at base temperature, at small fields (below the ESR resonance), with the YbRh$_{2}$Si$_{2}$ mounted. The stated power is the output power of the VNA, i.e.\ the actual power delivered to the resonator is somewhat lower due to damping of the microwave signal in the coaxial cables between VNA and resonator box. For the present study, the absolute values of applied power are not crucial, but the proper choice of the power for each resonance is important to achieve good signal-to-noice ratio without sample heating.
}
\label{FigPower}
\end{center}
\end{figure}
To assure this, we measured the dependence of the resonator quality factor on the applied microwave power for each resonator mode separately. We show exemplary power dependencies in Fig.\ \ref{FigPower}. 
Above a certain microwave power, the quality factor decreases substantially. We assign this suppression of $Q$ to microwave heating of the skin-depth region of the YbRh$_{2}$Si$_{2}$ sample. Our high-quality YbRh$_{2}$Si$_{2}$ (see Ref.\cite{krellner12a}) exhibits an increasing dc resistivity with increasing temperature down to temperatures well below our lowest temperature of 40 mK, and we observe a corresponding temperature dependence of our resonator $Q$ in zero magnetic field. Therefore, we assign both the temperature and power dependence of our resonator $Q$ to the YbRh$_{2}$Si$_{2}$ sample and not to the superconducting Nb coplanar line, whose superconducting critical temperature is much higher.  
For higher sample temperatures, we could increase the microwave power for improved signal-to-noise and faster measurements, while still assuring that no microwave heating occurred.
Data acquisition time of a single microwave spectrum (i.e.\ for one mode at one temperature and one magnetic field) was approximately one minute. For the lowest temperatures, the readout of the temperature sensor during this time period fluctuated by approximately 5 mK. 
The sample temperature that we assign to each particular ESR measurement is an average of the individual temperatures determined during the field sweep in the field range of the ESR absorption. For the lowest temperature, we estimate an error of approximately 3 mK for the experimental temperature of an individual ESR data point.

For each temperature of interest and at a sequence of static magnetic fields, we measure the transmitted microwave signal in a narrow spectral range around each of up to nine resonant frequencies per resonator. From this we determine the quality factor $Q$ of these resonance modes. This $Q$ depends on magnetic field in several ways: firstly, the microwave losses within the superconducting material of the resonator increase with magnetic field in a non-trivial fashion \cite{bothner12a}. Secondly, the conduction losses of the YbRh$_{2}$Si$_{2}$ sample, which are probed by the microwave electric fields in the resonator, depend on magnetic field as one moves through the phase diagram \cite{scheffler13a,parkkinen15a}. Thirdly, ESR is probed by the microwave magnetic fields when the static magnetic field fulfills the ESR condition. In this study we phenomenologically model the first two contributions as a continuous background that we then use to separate the ESR contribution \cite{JavaheriRahimDissertation}.

\section{Data analysis}
\subsection{Resonator modes}
We performed ESR measurements using three different coplanar resonators with different fundamental frequencies (1.50~GHz; 2.07~GHz; 2.65~GHz) and up to nine harmonics employed. Typical spectra are shown in Figs.\ \ref{FigSpecsOverview} and \ref{FigSpecsOverviewPD}. All the employed resonator frequencies and ESR resonance fields are compiled in the table T1.
\renewcommand{\thefigure}{S3}
\begin{figure}[tbph]
\begin{center}
\includegraphics[width=0.6\columnwidth]{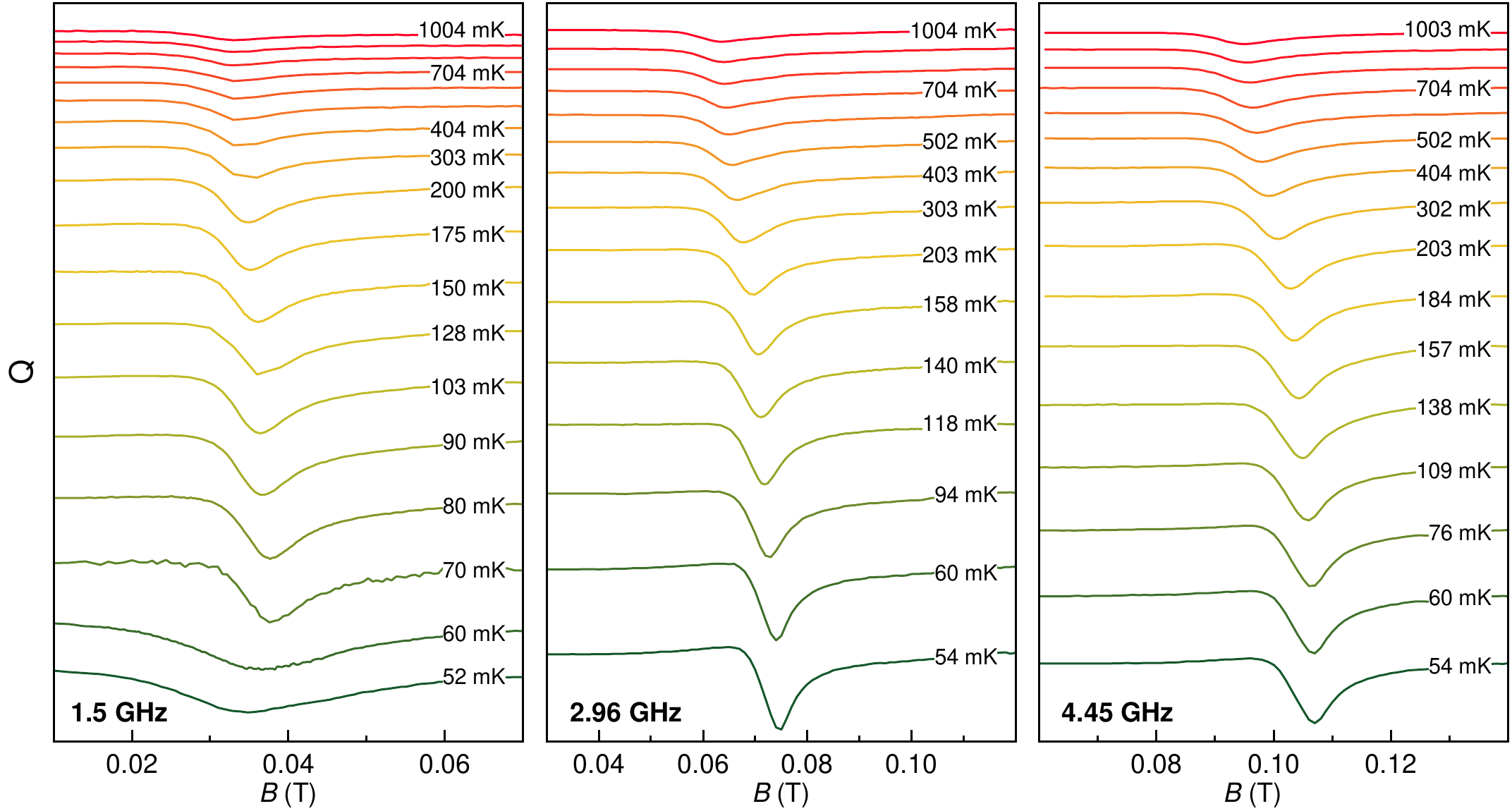}
\caption{
Overview of ESR spectra at selected temperatures for resonator R01 (harmonic 1, 2, 3) corresponding to the resonance fields shown as colored symbols in Fig.\ 1(a) and exemplary spectra in Fig.\ 1(b-d) of the main paper and the color-mapped spectra in the phase diagram in Fig.\ \ref{FigSpecsOverviewPD}(b).
}
\label{FigSpecsOverview}
\end{center}
\end{figure}
\renewcommand{\thefigure}{S4}
\begin{figure}[tbph]
\begin{center}
\includegraphics[width=0.4\columnwidth]{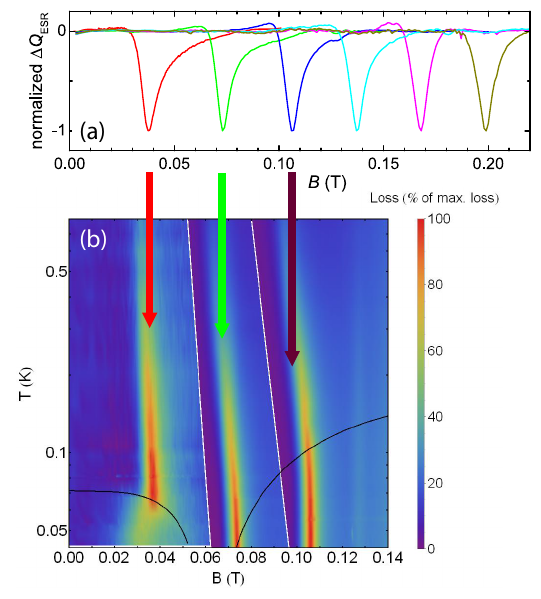}
\caption{(a) Examples of the first six ESR modes measured at $T=75$~mK by a coplanar resonator with fundamental frequency 1.5 GHz, i.e.\ ESR curves for the first six harmonics. We plot the magnetic field dependence of the quality factor $Q$, corrected with respect to the background magnetic field dependence; each curve is normalized to its minimum value. (b) Phase diagram with color-coded ESR contribution to resonator $Q$ for the first three modes. Similar to (a), the ESR data are background-corrected, but now quantifying increased absorption at the ESR as loss, with the ESR spectra of each of the three resonator modes normalized to the strongest ESR absorption, which occurs at different temperatures for the three modes. The black lines indicate boundaries of AFL and PFL regimes.}
\label{FigSpecsOverviewPD}
\end{center}
\end{figure}

\subsection{Fitting of ESR spectra}
We fitted the background-corrected ESR spectra with Lorentzian line shapes which are asymmetric because microwave dispersion effects occur due to the skin effect. More details on the ESR lineshape in YbRh$_{2}$Si$_{2}$ are reported in \cite{wykhoff07b}. Fig. \ref{FigSpecFit} shows examples of spectra together with Lorentzian line shapes (red solid lines) from which resonance field and linewidth were determined.
\renewcommand{\thefigure}{S5}
\begin{figure}[h]
\begin{center}
\includegraphics[width=0.4\columnwidth]{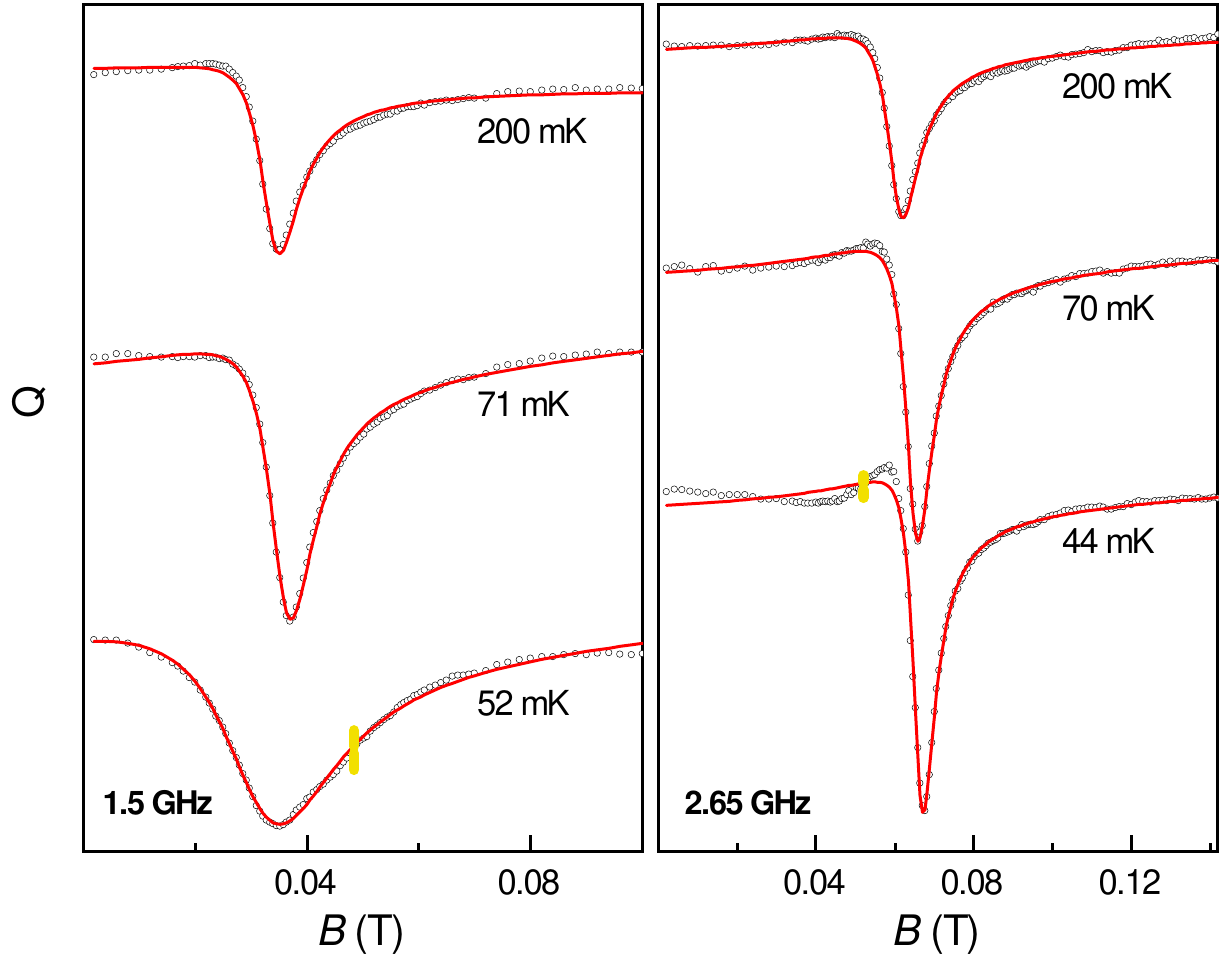}
\caption{
ESR spectra (symbols) and fitted Lorentzian line shape (solid lines) for resonator R01 (harmonic 1, 1.5 GHz) and resonator R03 (harmonic 1, 2.65 GHz). The yellow bars indicate boundaries between the AFL and PFL phases (see Fig.1). Note the deviations of the fits near these boundaries which can be understood in terms of the main-text discussion of the kink structures in Fig.\ 2(a).
}
\label{FigSpecFit}
\end{center}
\end{figure}
%
\renewcommand{\thetable}{T1}
\begin{table}[hb]
\begin{center}
\begin{tabular}{|c|c|c|c|c|}
\hline
Resonator label & Number of harmonic & Frequency& VNA output power & ESR field at 300 mK \\ \hline
R01	& 1&	1.50 GHz    & -40 dBm / -38 dBm &34 mT \\ \hline
R02	& 1&	2.07 GHz	& -41 dBm / -35 dBm &46 mT \\ \hline
R03 &	1&	2.65 GHz	& -47 dBm / -35 dBm &59 mT \\ \hline
R01 &	2&	2.96 GHz	& -50 dBm &67 mT \\ \hline
R02 &	2&	4.09 GHz	& -35 dBm &92 mT \\ \hline
R01 &	3&	4.45 GHz	& -47 dBm &100 mT\\ \hline
R03 &	2&	5.26 GHz	& -32 dBm &117 mT \\ \hline
R01 &	4&	5.88 GHz	& -50 dBm &130 mT \\ \hline
R02 &	3&	6.14 GHz	& -44 dBm &136 mT \\ \hline
R01 &	5&	7.34 GHz	& -47 dBm &162 mT \\ \hline
R03 &	3&	7.90 GHz	& -32 dBm &174 mT \\ \hline
R02 &	4&	8.13 GHz	& -38 dBm &179 mT \\ \hline
R01 &	6&	8.81 GHz	& -50 dBm &194 mT \\ \hline
R02 &	5&	10.13 GHz & -38 dBm &221 mT \\ \hline
R01 &	7&	10.21 GHz & -47 dBm &225 mT \\ \hline
R03 &	4&	10.52 GHz & -32 dBm &230 mT \\ \hline
R01 &	8&	11.86 GHz & -44 dBm &260 mT \\ \hline
R02 & 	6&	12.14 GHz & -50 dBm &262 mT \\ \hline
R01 &	9&	13.40 GHz & -38 dBm &293 mT \\ \hline

\end{tabular}
\caption{Overview of the resonator frequencies employed in this study. Three different resonators (labeled R01, R02, and R03) with fundamental frequencies 1.50 GHz, 2.07 GHz and 2.65 GHz were employed. The number of the harmonics and their frequencies are listed. The ESR resonance field at 300 mK is given as reference to relate ESR resonance frequencies and fields of the figures in the main text and to Fig.\ \ref{FigSpecsOverviewPD}.}
\end{center}
\label{Tab1}
\end{table}%

\clearpage

\section{Model for the magnetic resonance near the phase transition}
It was previously shown that in a paramagnetic state large contributions to the ESR linewidth due to the Kondo effect are canceled due to formation of a collective spin mode of the Yb ions and conduction electrons \cite{kochelaev09a}. Near the phase transition to the antiferromagnetic (AF) state the temperature dependence of the resonance magnetic field comes mainly from the reorientation of sublattices magnetizations relative to the external magnetic field. The renormalized constants of the Kondo and RKKY interactions can be considered in a phenomenological way, neglecting their temperature dependence. Having in mind an existence of strong ferromagnetic fluctuations in the planes perpendicular to the tetragonal symmetry axis, we use the following model: the Yb ions are coupled by a ferromagnetic (FM) exchange interaction $I_{1}$ in the planes and by an AF interaction $I_{2}$ between the planes. Choosing the x-axis along the symmetry axis of the crystal we use an effective spin Hamiltonian for the lowest Kramers doublet of the Yb ions and the conduction electrons in the following form:

\begin{eqnarray}
H&=&H^{pl}_{ex} + H^{bpl}_{ex} + H_{Z} + H_{se}; \\ \nonumber
H^{pl}_{ex}&=& -\frac{1}{2} \sum_{ij}^{nn} \left\{I_{1}\mathbf{S}_{i}\mathbf{S}_{j} - D_{1}S_{i}^{x}S_{j}^{x}\right\} ;\,
H^{bpl}_{ex}=\frac{1}{2} \sum_{ij}^{nnn} \left\{I_{2}\mathbf{S}_{i}\mathbf{S}_{j} - D_{2}S_{i}^{x}S_{j}^{x}\right\}.
\end{eqnarray}

\noindent Here, $D_{1}$ and $D_{2}$ represent the corresponding anisotropy of the exchange interactions, $I_{1}, I_{2}>0, nn$ (nearest neighbors) means a sum in the plane, $nnn$ (next nearest neighbors) means a sum between the planes. A similar model was used long ago by M.E. Lines to study antiferromagnetic ordering using the Green function method \cite{lines63a,lines64a}. The Zeeman energy and the interaction with conduction electrons can be represented by the Hamiltonians:

\begin{equation}
H_{Z}=-\mu_{\rm B}\sum_{j} \left[g_{\bot}\mathbf{HS}_{j}-(g_{\bot}-g_{\|})H_{x}S_{j}^{x}+g_{e}\mathbf{H}\boldsymbol{\sigma}_j\right];\;
H_{se}=\sum_{j} \left[ J\mathbf{S}\boldsymbol{\sigma}_{j} - J_{x}S_{j}^{x}\sigma_{j}^{x}\right].
\end{equation}

\noindent Here $\sigma_j^\alpha$ is the operator for the spin density of the conduction electrons at the Yb ion; $g_\bot{,g}_\parallel$ are $g$-factors perpendicular and along the tetragonal symmetry axis of the Yb ion, correspondingly; $g_e$ is the $g$-factor of the conduction electrons. It is important to note that the symmetry of all interactions is defined by the crystal electric field. Moreover, the anisotropy of the exchange constants for $H_{se}$ and $H_{ex}$ can be expressed via the $g$-factors similar to the Zeeman energy:
$J_{x}/J=g_{\bot}-g_{\|}, D_{1,2}^{2}/I_{1,2}=(g_{\bot}^{2}-g_{\|}^{2})/g_{\bot}^{2}$ (see Ref.\onlinecite{belov11a}).

\subsection{Molecular fields and the ESR frequency in the antiferromagnetic state.}

Since the anisotropy constants of the exchange interactions $D_{1}$ and $D_{2}$ should both be positive we expect an AF order of the ``easy plane'' type. In this case magnetizations of neighboring planes are coplanar and directed against each other. In the presence of an external magnetic field directed along the $z$-axis (perpendicular to the tetragonal axis of the crystal), the magnetizations of the neighboring planes rotate by some angle $\varphi$ toward the magnetic field. This angle can be found from the minimum of the free energy with respect to $\varphi$. It is convenient in this case to introduce for each sublattice a new set of coordinates with $z_{1}$-axis and $z_{2}$-axis along the $a$- and the $b$- magnetizations, correspondingly. The same transformation relates to the spin operators of conduction electrons for the corresponding planes. The new set of coordinates changes the Hamiltonian $H^{bpl}_{ex}$ and the Yb-ions part of $H_{Z}$ only.

\begin{eqnarray}
H^{bpl}_{ex}=\frac{1}{2}\sum_{ij}^{nnn}\left\{-I_2(S_{1i}^y S_{2j}^y + S_{1i}^z S_{2j}^z) \cos{2\varphi} + (I_2-D_2)S_{1i}^x S_{2j}^x \right\}; \\ \nonumber
H_{Z}=-\sum_{i(j)}\left\{g_{\bot}\mu_{\rm B}H(S_{1i}^z + S_{2j}^z)\sin{\varphi} + g_e\mu_{\rm B}H(\sigma_{1i}^z + \sigma_{2j}^z)\sin{\varphi}\right\}.
\end{eqnarray}

\noindent Here the index in the sum ($i$) relates to the first sublattice only and the index ($j$) to the second one. In the molecular field approximation, we have to put $\langle S_{a,b}^y\rangle = \langle \sigma_{a,b}^y\rangle =0$ and to introduce $\langle S_{a,b}^z\rangle = \bar S$ and $\langle \sigma_{a,b}^z\rangle = \bar\sigma$. The condition for the minimum of the free energy reduces to an equation $\langle \partial H/\partial\varphi\rangle=0$ which yields:

\begin{eqnarray}
\left\langle \frac{\partial H}{\partial\varphi}\right\rangle=2N\cos{\varphi}\left[ 2z_2I_2\bar S^2\sin{\varphi} - g_{\bot}\mu_{\rm B}H\bar S -  g_e\mu_{\rm B}H \bar\sigma \right]=0; \\ \nonumber
\text{solution 1:}\; \sin{\varphi}=\frac{g_{\bot}\mu_{\rm B}H\bar S +  g_e\mu_{\rm B}H \bar\sigma}{2n_2I_2\bar S^2},\;
\text{solution 2:}\; \varphi=\frac{\pi}{2}.
\end{eqnarray}

\noindent Here $N$ is the number of the Yb ions in the plane, $n_2$ is the number of their next nearest Yb ions (located in the two nearest planes). In the case $\bar\sigma=0$ this solution reduces to the result given by Lines \cite{lines64a}. The first solution is related to the AF state. It also yields the critical value of the external magnetic field that destructs the AF state. The second solution is related to the paramagnetic state, when the magnetizations of the two sublattices collapse into one directed along the external magnetic field. The equations for the average values $\bar S, \bar\sigma$ in the molecular field approximation are:

\begin{eqnarray}
S&=&\frac{1}{2}\tanh{\left\{\frac{1}{2k_{\rm B}T}\left[\bar S(z_1I_1+z_2I_2)-2z_2I_2\bar S\sin^2{\varphi}+g_{\bot}\mu_{\rm B}H\sin\varphi - \bar\sigma J \right] \right\}}; \\ \nonumber
\bar\sigma&=&\frac{1}{2}\rho_{\rm F}(g_e\mu_{\rm B}H\sin\varphi - \bar SJ).
\end{eqnarray}

\noindent Here $\rho_{\rm F}$ is the density of states at the Fermi surface of the conduction electrons. To find the frequencies of the ESR we have to write equations of motion for the total spins:

\begin{equation}
\mathbf{S}=\sum_{i\left(j\right)}{\left(\mathbf{S}_{1i}+\mathbf{S}_{2j}\right)\ \mathrm{,\ \ \ \ \ \ } \boldsymbol{\sigma}=\sum_{i\left(j\right)}{\left(\boldsymbol{\sigma}_{1i}+\boldsymbol{\sigma}_{2j}\right)\ }\ } 		
\end{equation}

\noindent After the Fourier transform in time and making linearization in the molecular field approximation equations of motion take the form:

\begin{eqnarray}
ES_x &=& i(g_{\bot}\mu_{\rm B}H\sin\varphi - \bar\sigma J)S_y + i\bar SJ\sigma_y; \\ \nonumber
ES_y &=& -i\left\{g_{\bot}\mu_{\rm B}H\sin\varphi + \bar S[\delta+z_2I_2(1+\cos2\varphi)]-\bar\sigma J\right\}S_x-i\bar S(J-J_x)\sigma_x; \\ \nonumber
E\sigma_x &=& i(g_e\mu_{\rm B}H\sin\varphi - \bar SJ)\sigma_y + i\bar\sigma JS_y ; \\ \nonumber
E\sigma_y &=& -i(g_e\mu_{\rm B}H\sin\varphi - \bar SJ)\sigma_x + i\bar\sigma(J-J_x)S_x .
\end{eqnarray}

\noindent Here $\delta=n_1D_1-n_2D_2$ ($n_1$ and $n_2$ are numbers of the (nn) and (nnn) ions correspondingly). In the following we consider the case of a very strong anisotropy ($g_{\|}\ll g_{\bot}$) which yields $J_x\approx J$. A condition for a nontrivial solution results in two frequencies; we are interested in a solution closest to the Yb ion Zeeman frequency. An approximate result for the case $\bar\sigma\bar S\ll1$ is the following:

\begin{equation}
E^2=\left\{g_{\bot}\mu_{\rm B}H\sin\varphi +  \bar S[\delta+n_2I_2(1+\cos2\varphi)] -\bar\sigma J \right\}(g_{\bot}\mu_{\rm B}H\sin\varphi - \bar\sigma J)
\end{equation}

\noindent Taking into account the first solution of Eq.~(4), we obtain for the effective $g$-factor of the antiferromagnetic resonance near the phase transition the following approximate result:

\begin{equation}
g_{\rm eff}^{\rm AF}=\frac{E_{\rm AF}}{\mu_{\rm B}H} \approx g_{s}\left( \frac{2\delta+4n_2I_2+\rho_{\rm F}J^2}{2n_2I_2}\right)^{1/2},\;
g_{s}=g_{\bot}-\rho_{\rm F}J
\end{equation}

\noindent The value of $n_2I_2$ can be estimated using the critical magnetic field $H_{c}$ for a destruction of the AF order at $T=0$. According to the first solution of the Eq. (4) for $\varphi=\pi/2$ we have

\begin{equation}
n_2I_2=\frac{g_{\bot}\bar S_{0}+g_{e}\bar\sigma_{0}}{2\bar S_{0}^2}\mu_{\rm B}H_{c} \approx \frac{g_{\bot}\mu_{\rm B}H_{c}}{2\bar S_{0}}.
\end{equation}

\noindent For an antiferromagnetic system $\bar S_{0}=\bar S(T=0)<S$ due to quantum fluctuations.  For an Yb ion with $S=\frac{1}{2}\mathrm{\ \ }{\bar{S}}_0=0.422$ (Ref.~\onlinecite{anderson52a}).

\subsection{Molecular fields and the ESR frequency in the paramagnetic state.}

In the paramagnetic state we have to use the second solution of Eq. (4): $\varphi=\pi/2$. The magnetizations of the two sublattices collapse into one that is directed along the external magnetic field. From Eq. (8) we obtain:

\begin{equation}
E_{\rm PM}^2 = (g_{\bot}\mu_{\rm B}H + \bar S_{\rm PM}\delta - \bar\sigma J)(g_{\bot}\mu_{\rm B}H - \bar\sigma J).
\end{equation}

\noindent The molecular fields can be found from the set of equations (5) with $\varphi=\pi/2$; their solution for $\bar\sigma\ll1$ gives:

\begin{equation}
\bar S_{\rm PM} = \frac{1}{2}\left[\frac{Tt_{0}}{T-(1-t_{0}^2)\theta}\right]; t_{0}=\tanh\frac{g_{s}\mu_{\rm B}H}{2k_{\rm B}T},
\theta=\frac{1}{4k_{\rm B}}\left(n_{1}I_{1}-n_{2}I_{2}+\frac{1}{2}\rho_{\rm F}J^2\right).
\end{equation}

\noindent At high temperatures the value $\theta$ plays the role of a Curie-Weiss temperature. For the field dependence of the effective $g$-factor we have from Eqs. (11,12):

\begin{equation}
g_{\rm eff}^{2}(H)=\left(\frac{E_{\rm PM}}{\mu_{\rm B}H}\right)^2 = \left(g_{s} + \bar S_{\rm PM}\frac{\rho_{\rm F}J^2}{2\mu_{\rm B}H}\right)\left(
g_{s} + \bar S_{\rm PM}\frac{2\delta+\rho_{\rm F}J^2}{2\mu_{\rm B}H}\right).
\end{equation}

\noindent For the temperature dependence of the $g$-factor we should have in mind that the resonance field itself depends on temperature too; a value of $g_{s}\mu_{\rm B}H$ can be found from Eq.(11):

\begin{equation}
g_{s}\mu_{\rm B}H=\sqrt{E_{\rm PM}^2+\frac{1}{4}(\bar S\delta)^2} - \frac{1}{2}\bar S(\delta + \rho_{\rm F}J^2).
\end{equation}

Taking into account $E_{\rm PM}=h\nu$ ($\nu$ and $h$ are the spectrometer frequency and the Planck constant, correspondingly), we obtain by iteration an approximate result:

\begin{equation}
g_{\rm eff}^{2}(T)=g_{s}^2\left(1+\bar S_{2}\frac{\rho_{\rm F}J^2}{2E_{2}}\right)\left(1+\bar S_{2}\frac{2\delta+\rho_{\rm F}J^2}{2E_{2}}\right);
\end{equation}

\noindent There is a choice for a starting value of Eq.~(14): at $T=\infty$ one should expect $\bar S=0$, $(g_{s}\mu_{\rm B}H)_{T=\infty}=h\nu$; at $T=0$ we can put $(g_{s}\mu_{\rm B}H)_{T=0}=E_{0}$ with $\bar S_{0}=0.422$ \cite{anderson52a}. It is reasonable near the phase transition to use the second case. Values of $E_{p}, \bar S_{p}$ with $p=0,1,2$ are given by

\begin{eqnarray}
E_{p}&=&\frac{1}{2}\left\{\sqrt{4(h\nu)^2+(\bar S_{p}\delta)^2} - \bar S_{p}(\delta+\rho_{\rm F}J^2)\right\}; \\ \nonumber
\bar S_{p}&=&\frac{1}{2}\left[\frac{Tt_{p}}{T-(1-t_{p}^2)\theta}\right];\; t_{1}=\tanh\frac{E_{0}}{2k_{\rm B}T},\; t_{2}=\tanh\frac{E_{1}}{2k_{\rm B}T}\;.
\end{eqnarray}

\subsection{Comparison of the model with experimental results.}

At first, we fit the field and temperature dependencies of the effective $g$-factor in the paramagnetic state.  Fig.\ref{FiggBT} represents the fit results with the formulas (13) and (15) for three temperatures $T=53,\ 125,\ 300,\ 500\ \mathrm{mK}$ and the two frequencies $\nu=1.5,\ 11.86\mathrm{\ GHz}$ with the same set of parameters. The obtained parameters are: $g_{\bot}=3.45, \delta=-0.15 \mathrm{\ K}, J=1.01 \mathrm{\ K}, \rho_{\rm F}=0.03 \mathrm{\ K}^{-1}$ (consistent with specific heat data at 0.2 K \cite{trovarelli00a}), $\theta=-0.01 \mathrm{\ K}$.  With decreasing temperature, the effective $g$-factor decreases and reaches the value $g_{\rm eff}^{\rm PM}=2.85$ for $\nu=1.5$~GHz at the critical temperature $T_{c}=57$~mK. Below the critical temperature the effective $g$-factor sharply increases. According to our model it experiences a jump to the value given by Eq.~(9). The value of the parameter $n_2I_2$ can be estimated from the critical magnetic field by Eq.~(10). Using the critical value of the field $B_c=60$~mT and parameters found above, we obtain $n_2I_2\approx165$~mK. Substituting this value and parameters found above into (9) we obtain $g_{\rm eff}^{\rm AF}\approx3.72$. One can see that the suggested model is able to explain a dramatic change of $g$-factor at the transition to the AF state.

An observed continuous increase of the $g$-factor (although very sharp) instead of a jump below the critical temperature can be related to an appearance of a mixed state with a volume of the AF state of the sample continuously increasing with decreasing of temperature. Such a mixed state is possible in the case of the phase transition of the first order, which actually happens in presence of the external magnetic field (since the symmetry of the system is not changing). With an appearance of the mixed state the antiferromagnetic resonance linewidth should start sharply increase in a non-homogeneous way. A sharp increase of the ESR linewidth below the critical temperature was actually observed experimentally.

\renewcommand{\thefigure}{S6}
\begin{figure}[htbp]
\begin{center}
\includegraphics[width=0.6\columnwidth]{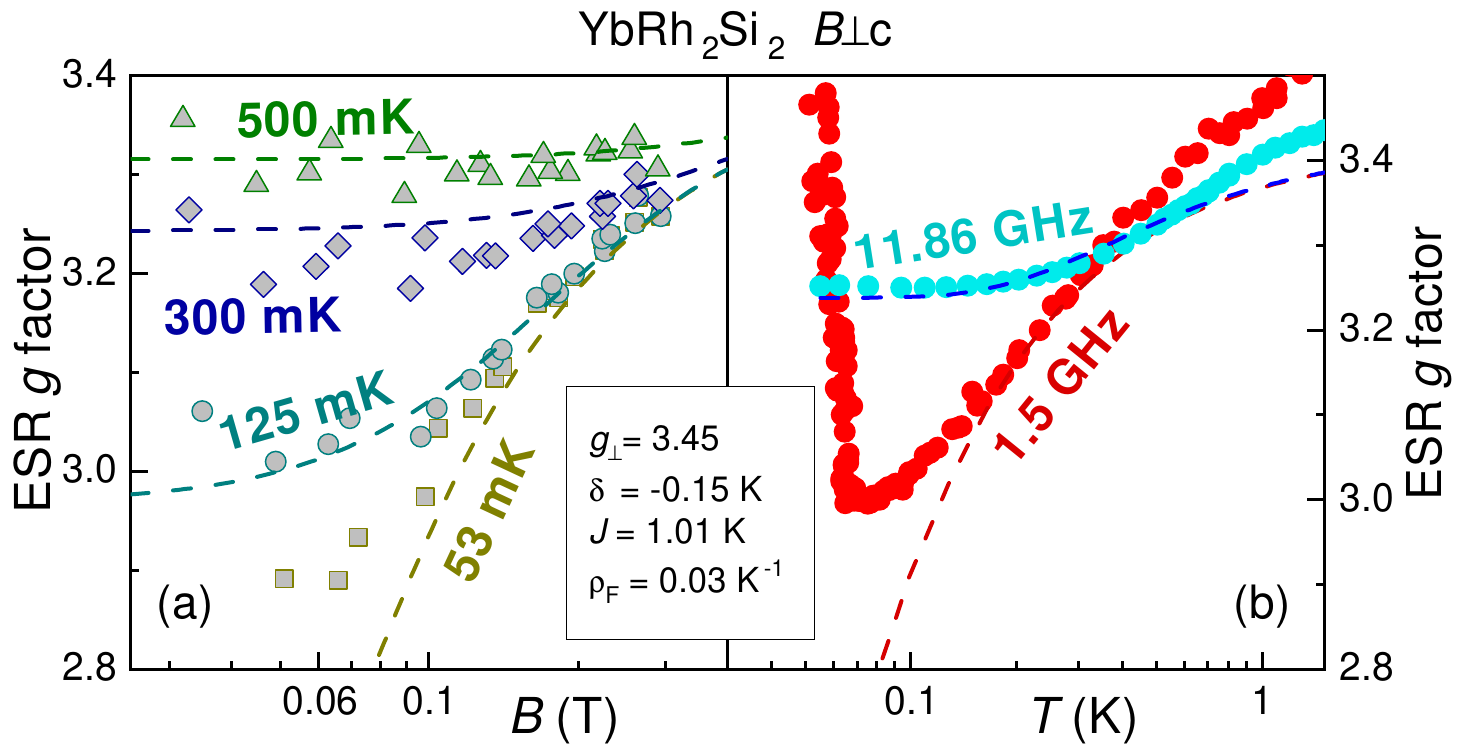}
\caption{The field- and temperature dependence of the effective $g$-factor for different temperatures. Dashed lines denote fitting with the formulas (13) and (15) with common set of parameters as listed in the inset.}
\label{FiggBT}
\end{center}
\end{figure}

\section{Comparison with a renormalized Zeeman splitting in a critical quasiparticle scenario}

In order to describe the peculiar properties of YbRh$_{2}$Si$_{2}$ at the field induced quantum critical point, W\"olfle et al.\ proposed a model based on spin flip scattering of critical quasiparticles \cite{wolfle15a}. In the course of setting up this model they estimated the renormalization of the Zeeman splitting from experimental thermodynamic data using a Fermi liquid approach. In this approach the ratio between the internal field $b$ and the external Field $B$ is essentially given by the Wilson ratio of magnetic susceptibility and specific heat \cite{wolfle15a}. Using published experimental data for the susceptibility and the specific heat, W\"olfle et al.\ calculated $b$ as a function of $B$ for a few temperatures in the interesting critical region of low temperatures and small fields, and showed the results in a $b$ versus $B$ plot in Fig.~1 of Ref.~\cite{wolfle15a}. As explained above, the $g$-factor measured in an ESR experiment reflects the ratio $b/B$. Therefore, we transformed Fig.~1 of Ref.~\cite{wolfle15a} to a plot $b/B$ as a function of $B$ (Fig.~\ref{FigbBB}), and compared this with the $g$-factor determined from our ESR experiments. The first observation is that this analysis based on a Fermi liquid approach and experimental data for the thermodynamic properties $C(T)$ and $M(B,T)$ results in huge renormalization factors of the Zeeman splitting. However, this is not reflected in the experimental ESR data, where the $g$-factors observed in this parameter range are reduced compared to the bare values determined from the crystalline electric field ground state. Therefore, in Fig.~\ref{FigbBB} we used different scales for $b/B$ and $g$, which allows to compare the relative $T$ and $B$ dependencies. The Fermi-liquid-thermodynamic approach indicates huge changes of the renormalization factor as a function of $T$ and $B$, which are also not seen in the experimental ESR results. Furthermore, the field dependencies do not match very well. Thus, the renormalization of the Zeeman splitting seen in the ESR experiment, which is a direct spectroscopic-microscopic technique, is very different from that deduced within an indirect Fermi liquid approach from thermodynamic properties. This illustrates the importance of the present ESR results in providing a direct, microscopic insight. A possible cause for the discrepancies between the huge renormalization factors deduced within a Fermi Liquid approach and those observed in ESR might be that the $4f$ electrons in YbRh$_{2}$Si$_{2}$ have a dual character, being partly itinerant, partly localized, a general feature of strongly correlated $f$-systems. Since the largest part of the magnetization likely originates from the localized part, while the largest part of the specific heat originates from the itinerant part, the ratio between susceptibility and specific heat shall be much larger than that corresponding to the Wilson ratio of the itinerant part solely.

\renewcommand{\thefigure}{S7}
\begin{figure}[htbp]
\begin{center}
\includegraphics[width=0.6\columnwidth]{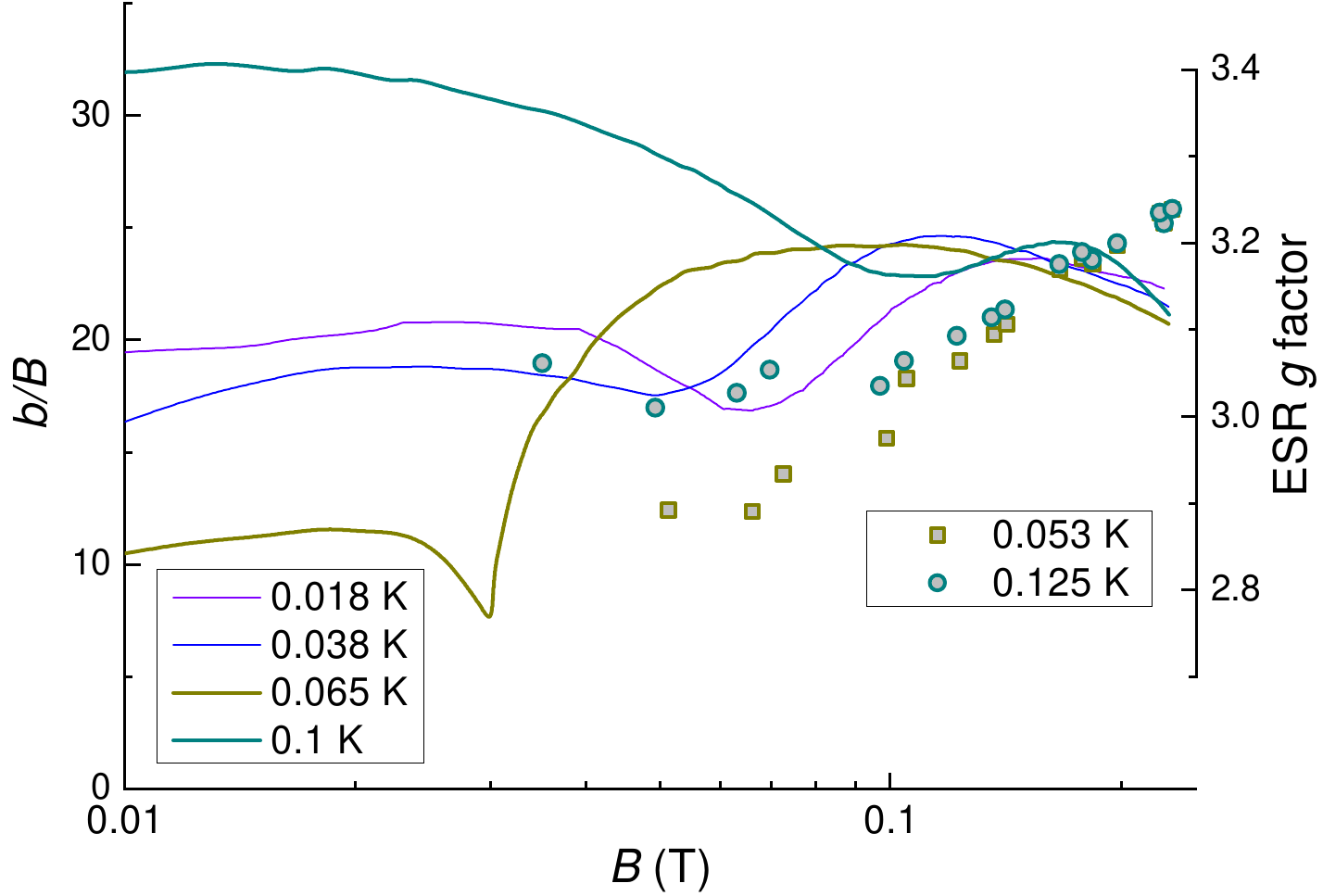}
\caption{The field-dependence of the renormalized Zeeman splitting (left axis, solid lines, from \cite{wolfle15a}) and measured effective $g$-factor (right axis, open symbols) for different temperatures.}
\label{FigbBB}
\end{center}
\end{figure}

\bibliography{YRSmKESRfinal}